\documentstyle[prl,twocolumn,aps]{revtex}
\begin{document}
\input psfig
\draft

\title{Strong Tunneling in the Single-Electron Box}

\author{J\"urgen K\"onig and Herbert Schoeller}

\address{Institut f\"ur Theoretische Festk\"orperphysik, Universit\"at
Karlsruhe, 76128 Karlsruhe, Germany}

\date{\today}

\maketitle

\begin{abstract}
We study strong tunneling (i.e. transmission $h/e^2R_T\gg 1$) in the 
single-electron box with many transverse modes at zero temperature. 
We develop a new renormalization group method which includes all charge states 
and requires no initial or final energy cutoff. 
We determine the ground-state energy, the average charge and the renormalized 
charging energy. 
The covered range for the coupling constant and the gate voltage is much 
increased in comparison to recent perturbative approaches, poor man scaling 
methods and Quantum Monte Carlo simulations. 
We reach the regime where Coulomb blockade become practically unobservable.
\end{abstract}
 
{\it Introduction.} 
Metallic nanostructures with junctions of low capacitance show quantization of 
electric charge and Coulomb blockade phenomena due to large charging energy 
\cite{Ave-Lik,Gra-Dev,Sch-Uebersicht}.
The simplest system in which one can see Coulomb blockade effects is the 
single-electron box.
Electrons enter or leave a small metallic island via a tunnel junction with
many transverse channels.
The junction is characterized by a resistance $R_T$ and a capacitance $C$.
The charge on the island is controlled by a capacitively (with $C_G$) coupled 
external voltage $V_G$.
In the ''weak tunneling'' regime, i.e., when the transmission 
${\cal T}=h/e^2 R_T$ of the barrier is much less than unity, the average 
charge of the box can be described within the ``orthodox theory''\cite{Ave-Lik} 
which treats tunneling in lowest order perturbation theory (golden rule). 
This means that the energy for charge $N$ on the island is given by the 
classical expression
\begin{equation}\label{energy}
E_N(n_x) = E_C (N-n_x)^2 \,,
\end{equation}
with $E_C = e^2/2(C+C_G)$ and $n_x=C_G V_G$. 
At zero temperature, the ground-state energy as function of $n_x$ is the 
periodic continuation of the parabola $E_0 (n_x)$ for $-1/2<n_x<1/2$. 
Furthermore, the average charge, given by the periodic continuation of
\begin{equation}\label{charge}
  \langle N \rangle (n_x) = n_x - {1\over 2E_C} {\partial E_0 \over 
\partial n_x}(n_x)\,,
\end{equation}
increases in steps at half-integer values of $n_x$. 

In this letter, we discuss the fundamental question how quantum fluctuations 
influence Coulomb blockade phenomena at zero temperature, i.e., we analyze the 
smearing of $E_0(n_x)$ and $\langle N \rangle(n_x)$ in the whole regime from weak
to strong tunneling up to ${\cal T}\sim 50$ (in recent experiments 
${\cal T}\sim 33$ was achieved \cite{Exp}). 
Another quantity of recent interest is the second derivative of the energy $E_0$,
\begin{equation}\label{charging energy}
E_C^*(n_x) = {1\over 2} {\partial^2 E_0 \over \partial n_x^2}(n_x) \,.
\end{equation}

Perturbative approaches \cite{Gra,Goe,KSS2} in $\alpha={\cal T}/(4\pi^2)$
up to $\alpha^3$ \cite{Goe} can describe $E_C^*(0)$ up to ${\cal T}\sim 15$.
However, within this approach, the range of allowed values for ${\cal T}$ 
decreases considerably for $n_x\rightarrow 1/2$, and the average charge 
turns out to be divergent at $n_x=1/2$. 

Therefore, nonperturbative approaches have been developed to describe the 
system near the degeneracy point of adjacent charge states either within a 
''non-crossing'' expansion scheme \cite{Gol-Zai} or using an effective two-state
system\cite{Mat,Fal-Schoen-Zim,SS,KSS1}.
But for ${\cal T}\gtrsim 1$ the broadening $\hbar/(R_T C) \sim {\cal T} E_C/\pi$
of the charge states is of the same order as the energy gap $E_C$, i.e., all 
charge states become important.
And even for weaker coupling the results from a renormalization group (RG) 
treatment\cite{Mat,Fal-Schoen-Zim} depend on both a high-energy and a low-energy 
cutoff. 
The first one, of the order of the charging energy, accounts for the neglected 
higher charge states. 
The second one is provided by the dominant low-energy scale, energy gap or 
temperature. 
As a result, quantitative statements depend on adjustable parameters.

For very strong coupling, ${\cal T}\gg 1$, instanton methods have been 
developed, which predict an exponential decrease of Coulomb blockade
phenomena: $E_C^*(0)/E_C\sim {\cal T}^\eta \exp{(-{\cal T}/2)}$,
\cite{Pan-Zai,Wan-Gra}. 
However, the exponent $\eta$ is still controversial.
Poor man scaling methods in $1/{\cal T}$ reveal a strong renormalization of
${\cal T}$ to smaller values \cite{Fal-Schoen-Zim,QMC Zwerger}. 
This analysis is performed in the phase representation for the topological 
sector with winding number zero. 
However, for decreasing ${\cal T}$ all winding numbers become important. 
Consequently, a quantitative matching to the two-charge state approximation is 
not possible \cite{GKSSZ}.

Even recent quantum Monte Carlo (QMC) simulations 
\cite{Goe,QMC Zwerger,QMC Wang,QMC Herrero} have led to different results for 
$E_C^*(0)$ in the regime ${\cal T}>15$.
Furthermore, the error bars and the computing time are very large for 
${\cal T}\gg 1$ and/or $n_x\rightarrow 1/2$ and extrapolation to zero 
temperature in order to describe ground state properties becomes difficult.
As a consequence, a large gap still remains between the small and large 
${\cal T}$-regions, the latter being even controversial.

It is the purpose of this letter to present an independent, real-time RG-method 
which covers self-consistently level-renormalization and -broadening together 
with vertex corrections, and includes all charge states without the need of 
initial or final energy cutoffs. 
In contrast to QMC simulations our theory works also at zero temperature.
For $E_C^*(0)$ we find good agreement with the QMC simulation of 
Refs.~\cite{Goe,QMC Zwerger}.
We also find agreement with the exponential behavior of $E_C^*(0)$ for 
${\cal T}\gtrsim 20$, as predicted by the controversial instanton results in 
Refs.~\cite{Pan-Zai,Wan-Gra} but with a different preexponent.
We calculate $\langle N \rangle(n_x)$ for all $n_x$ and predict that for
${\cal T} > 27$ charge quantization will be suppressed such that 
the deviation from a straight line is less than $1\%$. 
We find that the slope of $\langle N \rangle$ at $n_x=1/2$ is always infinite.

{\it Model.}
The system is modeled by the Hamiltonian $H=H_0+H_T$, in which
\begin{equation}
  H_0= \sum_{kn} \epsilon_{kn} a^\dagger_{kn}a_{kn} 
  + \sum_{qn} \epsilon_{qn} c^\dagger_{qn}c_{qn} 
  + \sum_N E_N |N\rangle \langle N| 
\end{equation}
describes the box without tunneling.
The operators $a^{(\dagger)}$ and $c^{(\dagger)}$ annihilate (create) electrons
in the lead and island electrode, respectively.
The tunneling coupling is given by
\begin{equation}
  H_T = \sum_N \sum_{\sigma=\pm} g_N^\sigma |N+\sigma \rangle \langle N| J_\sigma
\end{equation}
with $J_-=\sum_{kqn} T_{kq}^n a^\dagger_{kn} c_{qn}$ and $J_+=J_-^\dagger$.
The channel number $n=1,\ldots,N_{ch}$ accounts for both the transverse wave 
number and the spin and is conserved during a tunnel process.
We assume here a wide metallic junction with a large number of transverse 
channels. 
The matrix elements $T_{kq}^n$ are assumed to be independent of the states $k$ 
and $q$.
They are related to the tunnel resistance by $1/R_T = (2\pi e^2/\hbar) \sum_n
N_L(0) N_I(0) |T^n|^2$, where $N_{L/I}(0)$ are the density of states of the lead 
and the island.
The factor $g_N^\sigma$ describes the relative strength of the tunnel coupling.
At the beginning, we have $g_N^\sigma \equiv 1$ for all $N$ and $\sigma$.
But under the renormalization group procedure, the coupling will indeed depend
on the charge state $N$ and on $\sigma$.
We define $\alpha_N = \alpha g_N^+ g_{N+1}^-$.

{\it Renormalization group procedure.}
The invariant quantity under the RG is the S-matrix $S=T\exp (-i \int dt H(t))$.
The idea of the RG is to start with some high-energy cutoff, reducing it by 
integrating out the corresponding degrees of freedom, and end up with an 
effective low-energy theory.
All the energies and coupling constants will be renormalized, and,
simultaneously, new terms which are not present in the original Hamiltonian may 
be generated.
We will proceed in three steps.
(i) We introduce a time cutoff $t_c$.
(ii) The cutoff is increased $t_c\rightarrow t_c+dt_c$, thus flowing from 
$t_c^0$ to $t_c^f$.
The invariance condition for the S-matrix demands
\begin{equation}
  Te^{-i\int dt H(t)}\big|_{t_c} = Te^{-i\int dt H'(t)}\big|_{t_c+dt_c}
\end{equation}
for each RG step. 
(iii) We take $t_c^0\rightarrow 0$ and $t_c^f\rightarrow \infty$.

(i) The S-matrix reads in interaction picture 
$S=T\exp(-i\int dt H_0)T\exp(-i\int dt H_T(t)_I)$ in which $H_T(t)_I$ is the
tunnel part of the Hamiltonian in interaction picture with respect to $H_0$.
The expansion of the second exponential yields
\begin{equation}\label{S-matrix}
  S=Te^{-i\int dt H_0} \sum_{n=0}^\infty \,\,\,\,(-i)^n
  \!\!\!\!\!\!\int\limits_{t_1>t_2>\ldots>t_n} \!\!\!\!\!\!H_1\ldots H_n,
\end{equation}
where $H_i\equiv H_T(t_i)_I$.
Next, we perform a normal ordering of the fermion operators $c^{(\dagger)}$ and 
$a^{(\dagger)}$ involved in $H_i$,
\begin{eqnarray}
  H_1 &=& :H_1: 
  \label{normal1} \\
  H_1H_2 &=& :H_1H_2: + 
  H_1H_2 
  \begin{picture}(-20,11) 
    \put(-20,8){\line(0,1){3}} 
    \put(-20,11){\line(1,0){12}} 
    \put(-8,8){\line(0,1){3}}
  \end{picture}
  \label{normal2}  \\
  H_1H_2H_3 &=& :H_1H_2H_3: +
  H_1H_2
  \begin{picture}(-20,11) 
    \put(-20,8){\line(0,1){3}} 
    \put(-20,11){\line(1,0){12}} 
    \put(-8,8){\line(0,1){3}}
  \end{picture}
  \begin{picture}(20,11) 
  \end{picture}
  H_3 
  \nonumber \\ &&
  +
  H_1H_2H_3
  \begin{picture}(-33,11) 
    \put(-33,8){\line(0,1){3}} 
    \put(-33,11){\line(1,0){25}} 
    \put(-8,8){\line(0,1){3}}
  \end{picture}
  \begin{picture}(33,11) 
  \end{picture} +
  H_1H_2H_3  
  \begin{picture}(-20,11) 
    \put(-20,8){\line(0,1){3}} 
    \put(-20,11){\line(1,0){12}} 
    \put(-8,8){\line(0,1){3}}
  \end{picture}
  \begin{picture}(20,11) 
  \end{picture}
  \label{normal3}
\end{eqnarray}
where $:\ldots:$ stands for normal ordering and
$H_1H_2 
  \begin{picture}(-20,11) 
    \put(-20,8){\line(0,1){3}} 
    \put(-20,11){\line(1,0){12}} 
    \put(-8,8){\line(0,1){3}}
  \end{picture}
  \begin{picture}(20,11) 
  \end{picture}$
represents contractions of all the Fermi operators $c^{(\dagger)}$ and 
$a^{(\dagger)}$ in $H_1$ and $H_2$ with the relative time $t_1-t_2$.
In principle the two Fermi operators of $H_i$ may be contracted with Fermi 
operators of different $H_{i'}$'s.
For a large number of channels, $N_{ch}\rightarrow \infty$, however, these 
contributions are negligible.
The only combination of contractions which enter here is (at zero temperature)
\begin{equation}\label{contraction}
  \langle J_\pm(t_1)J_\mp(t_2)\rangle_{eq} 
  = - {\alpha \over (t_1-t_2)^2} + i \alpha \pi \delta'(t_1-t_2) .
\end{equation}
We now introduce a cutoff $t_c$ by multiplying the right hand side of 
Eq.~(\ref{contraction}) with $\Theta (|t_1-t_2|-t_c)$.
Consequently, via Eqs.~(\ref{S-matrix})-(\ref{normal3}), the S-matrix
is well-defined with a sharp cutoff $t_c$.

(ii) The increase of the cutoff $t_c'=t_c+dt_c$ leads to correction terms
$  H_1H_2 
  \begin{picture}(-20,11) 
    \put(-20,8){\line(0,1){3}} 
    \put(-20,11){\line(1,0){12}} 
    \put(-8,8){\line(0,1){3}}
    \put(-17.5,8.5){$\times$}
  \end{picture}
  \begin{picture}(20,11) 
  \end{picture} $,
$  H_1H_2H_3
  \begin{picture}(-33,11) 
    \put(-33,8){\line(0,1){3}} 
    \put(-33,11){\line(1,0){25}} 
    \put(-8,8){\line(0,1){3}}
    \put(-24,8.5){$\times$}
  \end{picture}
  \begin{picture}(33,11) 
  \end{picture} $,
etc. The cross indicates that the time difference is between $t_c$ and $t_c'$.
We interpret such many-time objects as clusters with {\it one} time argument
for the time ordering (we choose $t_2$).
It is possible to reexponentiate the sum of terms and correction terms.
The renormalized exponent reads
\begin{eqnarray}
  -i \int H_1' = && -i \int H_1 
  + (-i)^2 \! \int\limits_{t_1>t_2} H_1H_2 
  \begin{picture}(-20,11) 
    \put(-20,8){\line(0,1){3}} 
    \put(-20,11){\line(1,0){12}} 
    \put(-8,8){\line(0,1){3}}
    \put(-17.5,8.5){$\times$}
  \end{picture}
  \begin{picture}(20,11) 
  \end{picture}
  \nonumber \\ &&
  + \,(-i)^3 \!\! \int\limits_{t_1>t_2>t_3} \left\{ H_1H_2H_3 
  \begin{picture}(-33,11) 
    \put(-33,8){\line(0,1){3}} 
    \put(-33,11){\line(1,0){25}} 
    \put(-8,8){\line(0,1){3}}
    \put(-24,8.5){$\times$}
  \end{picture}
  \begin{picture}(33,11) 
  \end{picture}
  - H_2 H_1H_3
  \begin{picture}(-20,11) 
    \put(-20,8){\line(0,1){3}} 
    \put(-20,11){\line(1,0){12}} 
    \put(-8,8){\line(0,1){3}}
    \put(-17.5,8.5){$\times$}
  \end{picture}
  \begin{picture}(20,11) 
  \end{picture}
  \right\}
\end{eqnarray}
plus higher-order correction terms which we truncate.
This is the central equation since it determines the renormalization of all 
quantities.
The second term (propagator renormalization) can be absorbed in a change of the 
energies, while the third term (vertex corrections) renormalizes the coupling 
strength.

(iii) The limit $t_c^0\rightarrow 0$ is possible if we take into account all 
charge states.
There are no divergences like in a two-state approximation.
After performing the limit $t_c^f\rightarrow \infty$ we end up with an effective 
theory for the charge degrees of freedom without any contractions.

To present the results, we introduce the abbreviations $l=\ln (t_c/t_c^0)$, 
$\Delta_N=E_{N+1}-E_N$, $\bar{E}_N = E_N t_c$, and $\bar{\Delta}_N=\Delta_N t_c$.
The RG equations, then, are
\begin{eqnarray}\label{RG energy}
{d \bar{E}_N \over d l} &=& \bar{E}_N 
+ i \left( \alpha_{N-1} e^{i \bar{\Delta}_{N-1}} +\alpha_N e^{-i \bar{\Delta}_N} 
\right) \\
{d \bar{\Delta}_N \over d l} &=& \bar{\Delta}_N 
+ i \left( \alpha_N e^{i \bar{\Delta}_N} +\alpha_{N+1} e^{-i \bar{\Delta}_{N+1}} 
\nonumber \right. \\ && \left. \qquad \,\,\,\,\,\,
- \alpha_N e^{- i \bar{\Delta}_N} - \alpha_{N-1} e^{i \bar{\Delta}_{N-1}} 
\right) \label{RG delta} \\ 
{d \alpha_N \over d l} &=& \alpha_N \left\{ 
2 \alpha_{N+1} {e^{- i \bar{\Delta}_{N+1}}-e^{-i\bar{\Delta}_N}\over -i 
(\bar{\Delta}_{N+1}-\bar{\Delta}_N)}
\right. \nonumber \\ && \left. \,\,\,\,\,\,\,\,\,\,
+2 \alpha_{N-1} {e^{i \bar{\Delta}_N}-e^{i \bar{\Delta}_{N-1}}\over i 
(\bar{\Delta}_N-\bar{\Delta}_{N-1})} 
\right. \nonumber \\ && \left. \,\,\,\,\,\,\,\,\,\,
- \alpha_N e^{i\bar{\Delta}_N} -\alpha_{N+1} e^{-i \bar{\Delta}_{N+1}} 
\right. \nonumber \\ && \left. \,\,\,\,\,\,\,\,\,\,
- \alpha_N e^{- i \bar{\Delta}_N} - \alpha_{N-1} e^{i \bar{\Delta}_{N-1}} 
\right\} \, .
\label{RG alpha}
\end{eqnarray} 

{\it Discussion of the RG equations.}
We note the following:\\
-- All energies become complex, which is related to the broadening of the 
levels.\\
-- The coupling strength $\alpha_N$ and the difference $\Delta_{N+1}-\Delta_N$,
which are initially $\alpha$ and $2E_C$, respectively, evolve different for 
different charge states $N$.\\
-- The RG equations contain all terms from perturbation theory up to 
${\cal O}(\alpha^2)$.\\
-- Eqs.~(\ref{RG delta}) and (\ref{RG alpha}) include also all terms of the poor 
man's scaling. 
If one considers for $n_x\approx 1/2$ only two charge states $N=0,1$, and 
expands all exponentials 
$\exp (\pm i\bar{\Delta}_0)\approx 1 \pm i\bar{\Delta}_0$, Eqs.~(\ref{RG delta}) 
and (\ref{RG alpha}) reduce to the well-known formulas
$d \bar{\Delta}_0 /dl = \bar{\Delta}_0 (1-2\alpha_0)$ and
$d \alpha_0 /dl = -2\alpha_0^2$, with solutions 
$\Delta_0(l) / \Delta_0(0)=\alpha_0(l)/\alpha_0(0) = \left[ 1+2\alpha_0(0) l
\right]^{-1}$ with $\alpha_0(0) = \alpha$.

\begin{figure}
\centerline{\psfig{figure=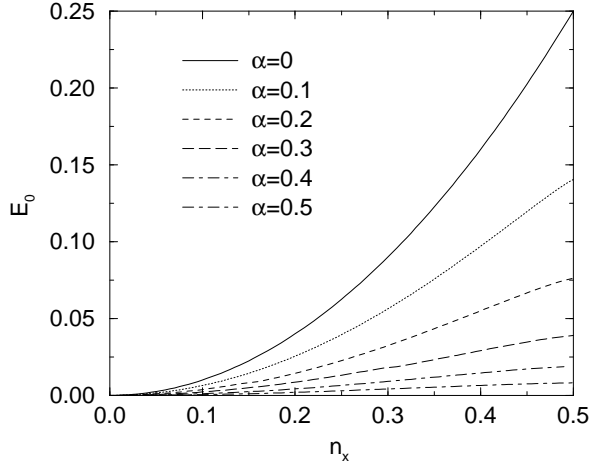,height=6.3cm}}
\caption{Ground-state energy as function of $n_x$.}
\label{fig ground state}
\end{figure}

\begin{figure}
\centerline{\psfig{figure=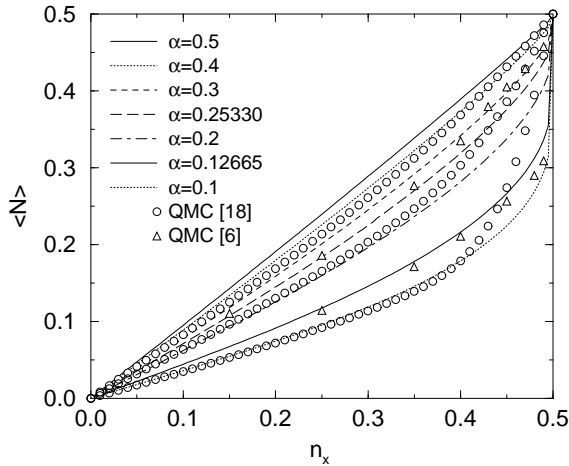,height=6.3cm}}
\caption[a]{Average charge as function of $n_x$.
The step at $n_x=1/2$ is rounded for increasing $\alpha$, approaching more and 
more a straight line.
The circles are QMC data from Ref.~\cite{QMC Herrero} for $\alpha=0.3, 0.2, 0.1$.
The triangles are QMC data from Ref.~\cite{Goe} for $\alpha=0.25330, 0.12665$.}
\label{fig average charge}
\end{figure}

\begin{figure}
\centerline{\psfig{figure=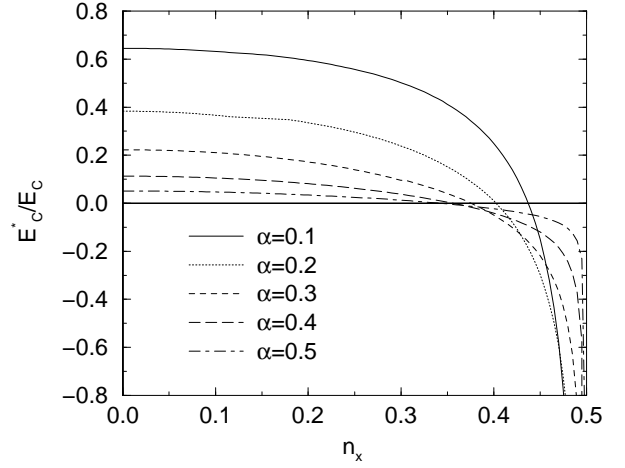,height=6.3cm}}
\caption{$E_C^*$ as a function of $n_x$.
For $n_x\rightarrow 1/2$ it diverges.}
\label{fig charging energy nx}
\end{figure}

\begin{figure}
\centerline{\psfig{figure=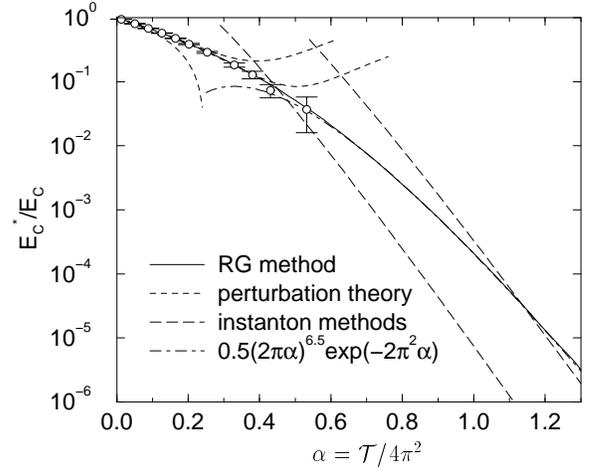,height=6.3cm}}
\caption[a]{$E_C^*$ as function of $\alpha$ for $n_x=0$.
Solid line: renormalization group method described in this letter.
Dashed lines: perturbation theory of order $\alpha$, $\alpha^2$, and $\alpha^3$ 
from Refs.~\cite{Gra,Goe}.
Long-dashed lines: large-$\alpha$ expansions from Refs.~\cite{Wan-Gra} 
(upper curve) and \cite{Pan-Zai} (lower curve).
Dot-dashed line: For $\alpha \gtrsim 0.5$ our data are consistent with
$0.5(2\pi\alpha)^{6.5}\exp(-2\pi^2\alpha)$.
Circles: QMC data from Ref.~\cite{Goe}.}
\label{fig charging energy log}
\end{figure}

\begin{figure}
\centerline{\psfig{figure=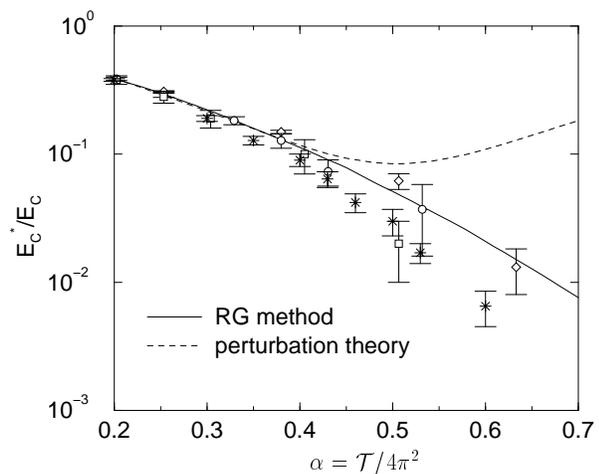,height=6.3cm}}
\caption[a]{$E_C^*$ as function of $\alpha$ for $n_x=0$.
Solid line: renormalization group method described in this letter.
Dashed lines: perturbation theory of order $\alpha^3$ \cite{Goe}.
QMC data from Ref.~\cite{QMC Wang} ($\diamond$), \cite{Goe} ($\circ$), 
\cite{QMC Zwerger} ($\Box$) and \cite{QMC Herrero} ($\ast$).} 
\label{fig QMC}
\end{figure}

{\it Results.}
In Figs.~\ref{fig ground state}-\ref{fig charging energy nx}, 
we show $E_0$, $\langle N \rangle$, and
$E_C^*$ as function of $n_x$ up to $\alpha=0.5$ (${\cal T}\sim 20$) and
compare with QMC data from Refs.~\cite{Goe,QMC Herrero}. 
(In the numerics we had to choose the initial cutoff $t_c^0$ much smaller than 
$0.1E_C^{-1}$. Here, we chose $0.01E_C^{-1}$.)
As expected we find a gradual flattening of the ground-state energy with 
increasing coupling and a rounding of the average charge towards a linear 
function.
The average charge $\langle N \rangle (n_x)$ agrees well with available 
QMC data for $\alpha \le 0.3$, except for $n_x\gtrsim 0.4$.
This deviation can be traced back to a finite temperature used in the QMC 
simulations \cite{com Herrero}. 
>From our results we conclude: 
(i) For ${\cal T}\sim 27$, the shape of $\langle N \rangle (n_x)$ is nearly 
indistinguishable from a straight line within an error of 1\%.
(ii) The slope of $\langle N \rangle$ at $n_x=1/2$ is infinite, i.e., there is 
always a very small crossover region around the degeneracy point showing a 
deviation from a straight line. 
This is explicitly shown in Fig.~\ref{fig charging energy nx} for the 
renormalized charging energy which diverges at $n_x=1/2$. 
However, this region is so small that it can hardly be identified experimentally.
(iii) The renormalized charging energy at $n_x=0$ decreases with increasing 
coupling. 
This reveals another indication for the washing out of charge quantization and 
is shown in more detail in Fig.~\ref{fig charging energy log} using a 
logarithmic scale for $E_C^*(0)$ as function of the coupling $\alpha$. 
Here we find very good agreement of our results with perturbation 
theory \cite{Gra,Goe} up to $\alpha \sim 0.4$ (${\cal T} \sim 15$) and
with the QMC data of Ref.~\cite{Goe}. Fig.~\ref{fig QMC} shows the
comparism to other QMC simulations. Including error bars, our 
data are consistent with the QMC data of Ref.~\cite{QMC Zwerger} 
(the authors report that the saturation for the largest ${\cal T}$ is still 
unclear \cite{com Hofstetter}), but are slightly lower than those of 
Ref.~\cite{QMC Wang}, and higher than those of Ref.~\cite{QMC Herrero}.
For ${\cal T} > 25$ no QMC data are available yet but the numerical solution
of the RG-equations (\ref{RG energy})-(\ref{RG alpha}) is stable until
${\cal T} \sim 50$. 
For ${\cal T} \gtrsim 20$ our data are consistent with the exponential behavior 
$E_C^*(0)/E_C \sim {\cal T}^\eta \exp{(-{\cal T}/2)}$ predicted by instanton
techniques \cite{Pan-Zai,Wan-Gra} but with $\eta=6.5$ rather than $\eta=2$ or 
$3$.

{\it Summary.}
We have investigated strong tunneling in the single-electron box with the 
help of a new, renormalization group method beyond the perturbative regime.
All charge states are included and, therefore, all results are cut-off 
independent.
We analyzed the observability of charging effects at zero temperature
by computing the average charge as function of the gate voltage.
We found a deviation of less than $1\%$ from a straight line for ${\cal T}>27$
and an infinite slope at the degeneracy point. Furthermore we compared
our results to contradicting QMC simulations and instanton techniques.

We acknowledge useful discussions with G. G\"oppert, H. Grabert, C.P. Herrero, 
G. Sch\"on, and A.D. Zaikin.
This work was supported by the ''Deutsche Forschungsgemeinschaft'' as part of 
''SFB 195''.

\end{document}